\documentstyle[preprint,aps,epsf]{revtex}
    \def\lsim{\raise0.3ex\hbox{$<$\kern-0.75em\raise-1.1ex\hbox{$\sim$}}}
\def\gsim{\raise0.3ex\hbox{$>$\kern-0.75em\raise-1.1ex\hbox{$\sim$}}}
\def\noi{\noindent}
\def\nn{\nonumber}
\def\bea{\begin{eqnarray}}  \def\eea{\end{eqnarray}}
\def\beq{\begin{equation}}   \def\eeq{\end{equation}}

\def\beeq{\begin{eqnarray}} \def\eeeq{\end{eqnarray}}

\begin{document}
\begin{center}
\vbox to 1 truecm {}
{\Large \bf J/$\psi$ suppression at $\sqrt{{\bf s}} =$ 200 GeV }\par
\vskip 3 truemm
 {\Large \bf in the comovers interaction model}
\vskip 1.5 truecm
{\bf A. Capella$^{\rm a)}$, E. G. Ferreiro$^{\rm b)}$}\\
\vskip 8 truemm

${\rm a)}$ Laboratoire de Physique Th\'eorique\footnote{Unit\'e Mixte de
Recherche UMR n$^{\circ}$ 8627 - CNRS}
\\ Universit\'e de Paris XI, B\^atiment 210,
F-91405 Orsay Cedex, France

\vskip 3 truemm

${\rm b)}$ Departamento de F{\'\i}sica de Part{\'\i}culas,
Universidade de Santiago de Compostela,\\ 15782 Santiago de Compostela,
Spain

\end{center}
\vskip -1 truecm
\begin{abstract}
The yield of $J/\psi$ per binary nucleon-nucleon
collision in $AuAu$ and $CuCu$ collisions at $\sqrt{s} = 200$~GeV is 
computed in
the framework of the dual parton model, supplemented with final state
interaction (comovers interaction). For the latter we use the same
value of the cross-section, $\sigma_{co} = 0.65$~mb, which describes
the anomalous $J/\psi$ suppression observed at CERN-SPS energies. 
Several possibilities for the value of the absorptive cross-section 
are considered.
Shadowing is introduced in both the comovers and the $J/\psi$ yields. 
A comparison with the results at CERN-SPS, including a prediction for 
$InIn$ collisions, is also presented.
  \end{abstract}

\vskip 0.75 truecm
\noi LPT-Orsay 05-23\par
\vskip -0.25cm
\noi April 2005
\newpage
\baselineskip = 20 pt
\pagestyle{plain}
\section{Introduction}
\hspace*{\parindent} $J/\psi$ production in proton-nucleus collisions
is suppressed with
respect to the characteristic $A^1$ scaling of lepton pair production
(Drell-Yan pairs). This suppression is generally interpreted as a
result of the multiple scattering of a pre-resonance $c-\overline{c}$
with the nucleons of the nucleus (nuclear absorption). In these
interactions, the $c-\overline{c}$ pair can transform into another
$c-\overline{c}$ pair with vanishing projection into $J/\psi$. The
corresponding cross-section $\sigma_{abs}$ is called absorptive
cross-section. This interaction is generally described in the
framework of a probabilistic Glauber model. However, at high energies,
the coherence length increases and the projectile interacts with the
nucleus as a whole. As a consequence the probabilistic Glauber formula
breaks down \cite{1r} \cite{2r} and, thus, the extrapolation from
CERN-SPS energies to RHIC ones ($\sqrt{s} = 200$~GeV) is not
straightforward.\par

The NA50 collaboration has observed \cite{3r} the existence of
anomalous $J/\psi$ suppression in $Pb-Pb$ collisions, i.e. the $J/\psi$
suppression in central $Pb-Pb$ collisions clearly exceeds the one
expected from nuclear absorption. Such a phenomenon was actually
predicted by Matsui and Satz \cite{4r} as a consequence of
deconfinement in a dense medium. It can also be described as a result
of final state interaction of the $c-\overline{c}$ pair with the dense
medium produced in the collision (comovers
interaction). The final results \cite{3r} of the NA50 collaboration can
be described using an effective cross-section $\sigma_{co} = 0.65$~mb
\cite{5r}. Since this is a low energy cross-section it is not expected
to change significantly in going from CERN-SPS to RHIC energies. Therefore the
prediction of the $J/\psi$ suppression due to comovers interaction at
RHIC seems to be quite  safe. However, at these energies it is
necessary to introduce shadowing corrections. These are
small at CERN-SPS energies. Moreover, they cancel to a large extend in
the ratio of $J/\psi$ over Drell-Yan pair production measured by NA50.
At RHIC energies, however, the Drell-Yan pair production is not
measured and the $J/\psi$ suppression is presented as a ratio of the
$J/\psi$ yield over the average number of binary nucleon-nucleon
collisions -- where the effect of shadowing is clearly present.

\section{The model}
\hspace*{\parindent} The ratio $R_{AB}^{J/\psi}(b)$ of the $J/\psi$
yield over the average
number of binary nucleon-nucleon collisions, $n(b)$, in $AB$ collisions
is given by
\beq
\label{1e}
R_{AB}^{J/\psi}(b) = {dN_{AB}^{J/\psi}(b)/dy  \over n(b)} =
dN_{pp}^{J/\psi}/dy  {\int d^2s\ \sigma_{AB}(b)\ n(b,s) \
S^{abs}(b,s)\ S^{co}(b,s) \over \int d^2s\ \sigma_{AB}(b)\ n(b,s)}\ .
\eeq

\noi Here $\sigma_{AB}(b) = 1 - \exp [- \sigma_{pp} AB T_{AB}(b)]$ where
$T_{AB}(b) = \int d^2sT_A(s) T_B(b-s)$ and $T_A(b)$ is the profile
function obtained from Wood-Saxon nuclear densities \cite{6r}, and
\beq
\label{2e}
n(b,s) = AB \ \sigma_{pp}\ T_A(s)\ T_B(b-s)/\sigma_{AB}(b) \ .
\eeq

\noi Upon integration of (\ref{2e}) over $d^2s$ we obtain the average
number $n(b)$ of binary nucleon-nucleon collisions at fixed impact
parameter $b$.\par

The factors $S_{abs}$ and $S_{co}$ in (\ref{1e}) are the $J/\psi$
survival probability due to nuclear absorption and comovers
interaction, respectively. \par

In writing eq. (\ref{1e}) we have assumed that the $J/\psi$ yield in
the absence of final state interaction (i.e. $S^{abs} = S^{co} = 1$)
scales with the number of binary nucleon-nucleon collisions. In this
case $R_{AB}^{J/\psi}$ coincides with the $J/\psi$ yield in $pp$
collisions.\\

{\bf a) \underbar{Comovers interaction}.} The survival probability
$S_{co}(b,s)$
of the $J/\psi$ due to comovers interaction is obtained by solving the
gain and loss differential equations which govern the final state
interactions with the co-moving medium \cite{7r}
\beq
\label{3e}
\tau {dN^{J/\psi}(b,s,y) \over d \tau} = - \sigma_{co}\
N^{J/\psi}(b,s,y)\ N^{co}(b,s,y)
\eeq

\noi where $N^{J/\psi}$ and $N^{co}$ are the densities (i.e. number per unit of
transverse surface) of $J/\psi$ and comovers (charged + neutral),
respectively. In eq. (\ref{3e}) we have neglected a gain term resulting
from the recombination of $c$ and $\overline{c}$ into $J/\psi$. This is
natural in our approach since  the cross-sections for recombination 
(gain) is expected to be substantially smaller than $\sigma_{co}$. 
The possibility of such a recombination, giving
sizable effects at RHIC energies, has been considered by several authors
\cite{8r}. It will be most interesting to see whether the data confirm
or reject such an effect.\par

In writing eq. (\ref{3e}) we have neglected transverse expansion and
assumed a dilution in time of the densities due to longitudinal motion
which leads to a $\tau^{-1}$ dependence on proper time $\tau$. Eq.
(\ref{3e}) can be solved analytically. The solution is invariant under
the change $\tau \to c \tau$. Thus, the result depends only on the
ratio $\tau_f/\tau_0$ of final over initial time. Using the inverse
proportionality between proper time and densities, we put
$\tau_f/\tau_0 = N^{co}(b,s,y)/N_{pp}(y)$, i.e. we assume that the
interaction stops when the densities have diluted, reaching the value 
of the $pp$ density at the same energy. At $\sqrt{s} =
200$~GeV and $y^* \sim 0$, $N_{pp}(0) = {3 \over 2}
(dN^{ch}/dy)^{pp}_{y^*=0}/\pi R_p^2 \sim 2.24$~fm$^{-2}$. Note the
increase in the $pp$ density form 1.15~fm$^{-2}$ at CERN-SPS to
2.24~fm$^{-2}$. Since the corresponding increase in the $AuAu$
densities is approximately the same, the average value of
$\tau_f/\tau_0$ is about the same at the two energies (of the order of
$5 \div 7$).\par

The solution of eq. (\ref{3e}) is given by \cite{5r}
\bea
\label{4e}
&&S^{co}(b,s) \equiv N^{J/\psi (final)}(b,s,y)/N^{J/\psi
(initial)}(b,s,y) \nn \\
&&= \exp \left [ - \sigma_{co}\ N^{co}(b,s,y) \ell n \left
(N^{co}(b,s,y)/N_{pp}(0)\right ) \right ]\ .
\eea
\vskip 5 truemm

{\bf b) \underbar{Comovers density in the dual parton model}.} The main
ingredient in order to compute the survival probability $S^{co}$ is
the comovers density $N^{co}$. Note that $N^{co}$ is the comovers
density at initial time $\tau_0$, i.e. the density produced in the
primary collision. It can be computed in the dual parton model
\cite{9r}. It can be seen from eq. (6.1) of ref. \cite{9r} that this
density is given by a linear combination of the average number of
participants and of binary nucleon-nucleon collisions, and for $A=B$
can be written as
\beq
\label{5e}
N_{NS}^{co}(b,s,y) = {3 \over 2} \ {dN_{NS}^{ch} \over dy}(b,s,y) =
{3 \over 2} \left [ C_1(b)\ n_A(b,s) + C_2(b)\ n(b,s)\right ]
\eeq

\noi where
\beq
\label{6e}
\left . n_A(b,s) = A\ T_A(s) \left [ 1 - \exp \left ( - \sigma_{pp} —
B \ T_B(b-s)\right )\right ]\right / \sigma_{AB}(b)
\eeq

\noi and $n(b,s)$ is given by eq. (\ref{2e}). The factor 3/2 takes care
of the neutrals. The coefficients $C_1(b)$ and $C_2(b)$ are obtained
from string multiplicities which are computed in DPM as a convolution
of momentum distributions functions and fragmentation functions. These
functions are universal, i.e. the same for all hadronic and nuclear
processes. Thus, we use the same expressions as at CERN energies. For
details see \cite{10r}. The numerical values of $C_1(b)$ and $C_2(b)$
in $AuAu$ and $CuCu$ collisions at $\sqrt{s} = 200$ computed in the 
rapidity interval
$|y^*| < 0.35$ for various values of $b$ and per unit rapidity are
given in Table 1. We also give in this table the corresponding 
values for $PbPb$ and $InIn$ at $p_{lab} = 158$~GeV/c.\par

We see from Table 1 that $C_2$ is significantly larger than $C_1$
at RHIC energies. Thus, DPM leads to
multiplicities which have a behaviour closer to a scaling with the
number of binary collisions rather than to a scaling with the number of
participants. Actually, with increasing energies the ratio $C_2/C_1$
increases and one obtains a scaling in the number of binary collisions.
This is a general property of Gribov's Reggeon Field Theory which is
known as AGK cancellation \cite{11r} -- analogous to the factorization
theorem in perturbative QCD and valid for soft collisions in the
absence of triple Pomeron diagrams. It is well known that this
behaviour is inconsistent with data which show a much smaller increase
with centrality. As discussed in detail in \cite{10r} such a
discrepancy is due to shadowing which is important at RHIC energies and
has not been taken into account in eq. (\ref{5e}). This is precisely
the meaning of label NS (no shadowing) in this equation.\\

{\bf c) \underbar{Shadowing}.} Following ref. \cite{10r,12r,13r}
shadowing corrections are computed, without free parameters, in terms
of the measured value of the diffractive cross-section. Indeed, in the
framework of Gribov's Reggeon Field Theory, the same triple Pomeron
diagrams which describe high mass diffraction are
responsible for the shadowing corrections. While the contribution of
the triple Pomeron diagram to high mass diffraction is positive, its
contribution to the total cross-section is negative, due to the
presence of $s$-channel discontinuities which correspond to
interference terms. Thus, the triple Pomeron diagrams (with triple
Pomeron coupling determined from high-mass diffraction data) produce a
decrease of the charged yield as given by eqs. (\ref{5e})-(\ref{7e}),
thereby violating the AGK cancellation. In $AB$ collisions this
reduction is given\footnote{A more sophisticated calculation using
other triple Regge diagrams, with parameters constrained from HERA data
can be found in ref. \cite{13r}. The results, however, are very similar
to the ones obtained from eqs. (\ref{7e})-(\ref{8e}).} by
\cite{10r,12r}
\beq
\label{7e}
S_{ch}^h(b,s,y) = {1 \over 1 + A \ F_h(y)\ T_A(s)} \ {1 \over 1 + B\
F_h(y)\ T_B(b-s)}\ .
\eeq

\noi Here the function $F(y)$ is given by the integral of the ratio of
the triple Pomeron cross-section $d^2\sigma^{PPP}/dY dt$ at $t= 0$ over
the single Pomeron exchange cross-section $\sigma_{P}$~:
\beq
\label{8e}
\left . F_h(y) = 4 \pi \int_{Y_{min}}^{Y_{max}} dY {1 \over
\sigma_P}\ {d^2 \sigma^{PPP} \over dY dt}\right |_{t=0} = C \left [
\exp \left (\Delta Y_{max}\right ) - \exp \left ( \Delta
Y_{min}\right )\right ]
\eeq

\noi where $Y_{min} = \ell n (R_Am_N/\sqrt{3})$, $\Delta = 0.13$ and $C
= 0.31$~fm$^2$. The value of $Y_{max}$ depends on the rapidity of the
produced particle. For $y^* = 0$ we have $Y_{max} = 1/2 \ell n
(s/m_T^2)$ where $m_T$ is the transverse mass of the produced particle.
For charged particles we use $m_T = 0.4$~GeV and for a $J/\psi$ $m_T =
3.1$~GeV.\par

It has been shown in \cite{10r} that with the shadowing resulting from
eqs. (\ref{7e})-(\ref{8e}) a good description of the centrality
dependence of charged multiplicities at mid-rapidities is obtained at 
RHIC energies ($\sqrt{s} = 130$ and 200~GeV).
More precisely one has
\beq
\label{9e}
N^{co}(b,s,y) = N_{NS}^{co}(b,s,y) \ S_{sh}^{ch}(b,s,y)
\eeq

\noi where the two factors in the r.h.s. are given by eqs. (\ref{5e})
and (\ref{7e}), respectively. \par

With this expression of the density of
comovers we can compute the $J/\psi$ survival probability $S^{co}$, eq.
(\ref{4e}). The $J/\psi$ suppression $R_{AB}^{J/\psi}$ is given by eq.
(\ref{1e}) with the following replacement in its numerator
\beq
\label{10e}
n(b,s) \to n(b,s)\ S_{sh}^{J/\psi}(b,s,y) \ .
\eeq

\noi Indeed, as discussed above, in writing the numerator of eq.
(\ref{1e}) we have assumed that the $J/\psi$ yield in the absence of
final state interaction ($S^{abs} = S^{co} = 0$) scales with the number
of binary collisions. This is only true when shadowing is neglected.
The effects of shadowing on the $J/\psi$ yield are introduced with 
the replacement
(\ref{10e}) in the numerator of eq. (\ref{1e}).\\

{\bf d) \underbar{Nuclear absorption}.} The formula for nuclear
absorption used in the literature is obtained in a probabilistic
Glauber model. One has
\beq
\label{11e}
S^{abs}(b,s) = {[1 - \exp (-A \ T_A(s)\ \sigma_{abs} ) ] [1 - \exp
(-B\ T_B (b-s) \sigma_{abs})] \over \sigma_{abs}^2 \ AB\ T_A(s)\
T_B(b-s)}\ .
\eeq

\noi As discussed in the Introduction, this formula breaks down at
high-energy due to the increase of the coherence length \cite{1r}
\cite{2r}. In the limit of $s\to \infty$, the relevant equation is
quite simple. The change consists in the replacement
\beq
\label{12e}
\left (1/\sigma_{abs}\right ) \left [ 1 - \exp \left ( -
\sigma_{abs}\ A\ T_A(b) \right ) \right ] \Rightarrow A\ T_A(b) \exp
\left [ - {1 \over 2} \sigma_{c\overline{c}-N}\ A\ T_A(b)\right ]
\eeq

\noi in each of the two factors in the numerator of (\ref{11e}).
The corresponding formula at finite energies which interpolates between
(\ref{1e}) and (\ref{2e}) has also been derived \cite{1r}. The change
in going from (\ref{11e}) to (\ref{12e}) is twofold. There is a change
in the form of the expression and, moreover, $\sigma_{abs}$ has been
replaced by the total $c\overline{c}-N$ cross-section
$\sigma_{c\overline{c}-N}$. If $\sigma_{c\overline{c}-N} \sim
\sigma_{abs}$ the change from low energies to asymptotic ones is small.
Indeed the two expressions coincide at the first and second order in
the development of the exponential, and since $\sigma_{abs}$ is
small, the low energy result will not be significantly changed.
However, if $\sigma_{c\overline{c}-N}$ is significantly larger than its
absorptive part, $\sigma_{abs}$, the $J/\psi$ suppression due to final
state interaction within the nucleus will be larger at high energies.
The latter possibility seems to be ruled out by preliminary data
\cite{14r} on $dAu$ collisions which show a rather small suppression at
mid-rapidities.\par

In the next section we present the calculation of $J/\psi$ suppression
in $AuAu$ and $CuCu$ collisions using eq. (\ref{11e}) with the value 
$\sigma_{abs}
= 4.5$ obtained from the $pA$ data at CERN-SPS. With these values
of $\sigma_{abs}$ the results obtained with eqs. (\ref{1e}) and
(\ref{2e}) are practically the same. We also present the results
obtained with smaller values of $\sigma_{abs}$ ($\sigma_{abs} = 0$, 1 
mb and 3 mb).

\section{Numerical results}
\hspace*{\parindent} We compute first the inclusive charged particle
multiplicity given by
\beq
\label{13e}
{dN^{ch} \over dy} (b,y) = \int d^2s {dN_{NS}^{ch} \over dy} (b,s,y)
\ S_{sh}^{ch} (b,s, y)\ .
\eeq

\noi At mid-rapidities, this quantity can be computed at various
centralities using the coefficients $C_1(b)$ and $C_2(b)$ in Table~1
and eq. (\ref{7e}).
The calculations for $AuAu$ collisions at mid-rapidities are shown in 
Fig.~1 and compared
with experiment
\cite{15r}. As predicted in \cite{10r} a reasonable description of the data
is obtained. An increase by a factor 1.13 between $\sqrt{s} = 130$~GeV
and $\sqrt{s} = 200$~GeV for central collision was also predicted in
\cite{10r} -- in
agreement with present data.\par

In Fig.~2 we compare the result of our calculations of 
$R_{AB}^{J/\psi}(b)$ in eq. (\ref{1e}) for different systems~: 
$PbPb$\footnote{The results for $PbPb$ are identical to those in the 
first paper of \cite{5r}, except that in \cite{5r} the ratio $J/\psi$ 
over $DY$ was plotted versus $E_T$ (the energy deposited in the NA50 
calorimeter). Moreover, for large $E_T$ (beyond the ``knee'' of the 
$E_T$ distribution) the effect of the fluctuation in the comovers 
multiplicity was included. This is not the case in Fig.~2 since, in a 
plot versus $N_{part}$, such a situation does not arise.} and $InIn$ 
at CERN-SPS ($p_{lab} = 158$~GeV/c) and $AuAu$ and $CuCu$ at 
$\sqrt{s} = 200$~GeV. In all cases the normalization is arbitrary but 
the same for all. It corresponds to taking $dN_{pp}^{J/\psi}/dy= 1$ 
in eq. (\ref{1e}). Also in all cases we use  $\sigma_{co} = 0.65$~mb 
and $\sigma_{abs} = 4.5$~mb. An important feature of our results is 
that, at a given energy, the results for the lighter systems are 
rather close to the ones for the heavier ones, at the same values of 
$N_{part}$. We also see that the $J/\psi$ suppression is much larger 
at RHIC energies and reaches a factor 10 for central $AuAu$ 
collisions.\par

In Fig. 3 we present again the result of our calculation of 
$R_{AuAu}^{J/\psi}(b)$ in eq. (\ref{1e}), with the normalization 
given by the measured value of $dN_{pp}^{J/\psi}/dy$ at $\sqrt{s} = 
200$~GeV \cite{16r}. All curves are obtained with $\sigma_{co} = 
0.65$~mb and different values of $\sigma_{abs}$ ($\sigma_{abs} = 0$, 
1mb, 3 mb and 4.5 mb). The suppression for central collisions varies 
between a factor of 6 for $\sigma_{abs} = 0$ and a factor of 10 for 
$\sigma_{abs} = 4.5$~mb. Even in the former case the suppression is 
twice as large as the one obtained in a QCD based nuclear absorption 
model \cite{17r}.\par

The results in Fig.~2 for $PbPb$ and $InIn$ has been obtained without 
including shadowing. Using eq. (\ref{7e}) it turns out that, at 
CERN-SPS, the shadowing on the $J/\psi$ is negligibly small. However, 
for the comovers $(m_T = 0.4$~GeV), its effect is of the order of 
15~\%. Here the values of $Y_{max}$ and $Y_{min}$ in eq. (\ref{8e}) 
are quite close to each other and our equations (and in particular 
the expression of $Y_{min}$) are not accurate enough for a reliable 
calculation. If, however, an effect of shadowing of the order of 
15~\% is present, the values of $C_1$ and $C_2$ in Table 1 should be 
increased by the same amount in order to restore agreement \cite{10r} 
with the experimental values of the charged multiplicities in $PbPb$ 
at $p_{lab} = 158$~GeV/c. This, in turn, would result in a larger 
$J/\psi$ suppression at RHIC. The maximal effect occurs in the case 
$\sigma_{abs} = 0$ and is of the order of 20~\% for central $AuAu$ 
collisions.

\section{Conclusions}
\hspace*{\parindent} In a comovers interaction framework we have 
computed the yield of $J/\psi$ per binary nucleon nucleon collision 
versus the number of participants in $PbPb$ and $InIn$ collisions at 
CERN-SPS ($p_{lab} = 158$ GeV/c) and in $AuAu$ and $CuCu$ at 
$\sqrt{s}=200$~GeV. At RHIC energies shadowing corrections to both 
the $J/\psi$ and the comovers multiplicities are very important and 
have been included in the calculations. We have found that, at a 
given energy, the $J/\psi$ suppression for the lighter and heavier 
systems are similar, at the same value of $N_{part}$. We have also found that the $J/\psi$ suppression at RHIC is significantly larger than 
at SPS. For central $AuAu$ collisions it reaches a factor of 10 for 
$\sigma_{abs} = 4.5$~mb and a factor 6 for $\sigma_{abs} = 0$. The 
value of $\sigma_{abs}$ has to be determined from the $dAu$ data. 
Preliminary results \cite{16r} favor a rather small value, 
$\sigma_{ab} \approx 1$~mb.\par

We have argued that these values could be underestimated by about 
20~\%. Experimental values of the $J/\psi$ suppression significantly 
smaller that the one in Fig.~3 would not be consistent with the 
comovers interaction model, at least in its present formulation.\par

Finally, an important difference between the $J/\psi$ suppression 
pattern in a comovers interaction model and in a deconfining scenario 
is that, in the former case, the anomalous supression sets in 
smoothly from peripheral to central collisions -- rather than in a 
sudden way when the deconfining threshold is reached. The NA50 
results have not allowed to disentangle these two possibilities. 
However, at RHIC energies, the relative contribution of the comovers 
is strongly enhanced in our approach, and a clear cut answer to this 
important issue should be obtained.

\noi {\Large \bf Acknowledgments} \par\nobreak
It is a pleasure to thank N. Armesto for interesting discussions and
helpful suggestions. E. G. Ferreiro also thanks the
Service de Physique Th\'eorique, CEA, Saclay, for
hospitality 
during the completion of this work.

\begin{table}[htbp]
\begin{center}
\begin{tabular}{|c|c|c|c|c|c|c|c|c|}
b & $C_1$ AuAu & $C_2$ AuAu & $C_1$ CuCu & $C_2$ CuCu
& $C_1$ PbPb & $C_2$ PbPb & $C_1$ InIn & $C_2$ InIn\\
\hline
0. &    1.0274 &1.7183 & 1.0330 & 1.8196  &  0.7102 & 0.3975 & 0.7480 & 0.4312 \\
1. &    1.0276 &1.7206 & 1.0334 & 1.8239  &  0.7115 & 0.3987 & 0.7485 & 0.4317 \\
2. &    1.0278 &1.7228 & 1.0338 & 1.8320  &  0.7152 & 0.4020 & 0.7527 & 0.4357 \\
3. &    1.0286 &1.7340 & 1.0342 & 1.8437  &  0.7208 & 0.4070 & 0.7599 & 0.4428 \\
4. &    1.0293 &1.7448 & 1.0347 & 1.8592  &  0.7283 & 0.4136 & 0.7696 & 0.4526 \\
5. &    1.0302 &1.7574 & 1.0352 & 1.8787  &  0.7376 & 0.4218 & 0.7810 & 0.4646 \\
6. &    1.0310 &1.7722 & 1.0357 & 1.9014  &  0.7488 & 0.4320 & 0.7945 & 0.4793 \\
7. &    1.0320 &1.7908 & 1.0361 & 1.9258  &  0.7617 & 0.4445 & 0.8112 & 0.4985 \\
8. &    1.0330 &1.8121 & 1.0364 & 1.9505  &  0.7764 & 0.4597 & 0.8290 & 0.5198 \\
9. &    1.0340 &1.8374 & 1.0364 & 1.9754  &  0.7929 & 0.4776 & 0.8475 & 0.5430 \\
10.&    1.0349 &1.8665 & 1.0363 & 2.0006  &  0.8112 & 0.4985 & 0.8664 & 0.5681 \\
11.&    1.0357 &1.8990 & 1.0360 & 2.0259  &  0.8308 & 0.5220 & 0.8855 & 0.5949 \\
12.&    1.0362 &1.9308 & 1.0356 & 2.0515  &  0.8503 & 0.5466 & 0.9046 & 0.6235 \\
13.&    1.0364 &1.9580 & 1.0349 & 2.0772  &  0.8673 & 0.5698 & 0.9233 & 0.6536 \\
\end{tabular}
\end{center}
\caption{Values
of $C_1$ and $C_2$ in eq.
(\ref{5e}) as a function of the impact parameter $b$. The second and
third columns correspond to $AuAu$ collisions and the forth and fifth
to $CuCu$ collisions both at $\sqrt{s} = 200$~GeV. The values,
calculated in the range $-0.35 < y^*  < 0.35$, are given per unit
rapidity. The following columns refer to $PbPb$ and $InIn$ at
$p_{lab} = 158$~GeV/c and are computed in the rapidity range of the
NA50 dimuon trigger $0 < y^* < 1$.}
\end{table}



\begin{figure}
\centering\leavevmode
\vskip 3.5cm
\epsfxsize=6in\epsfysize=6in\epsffile{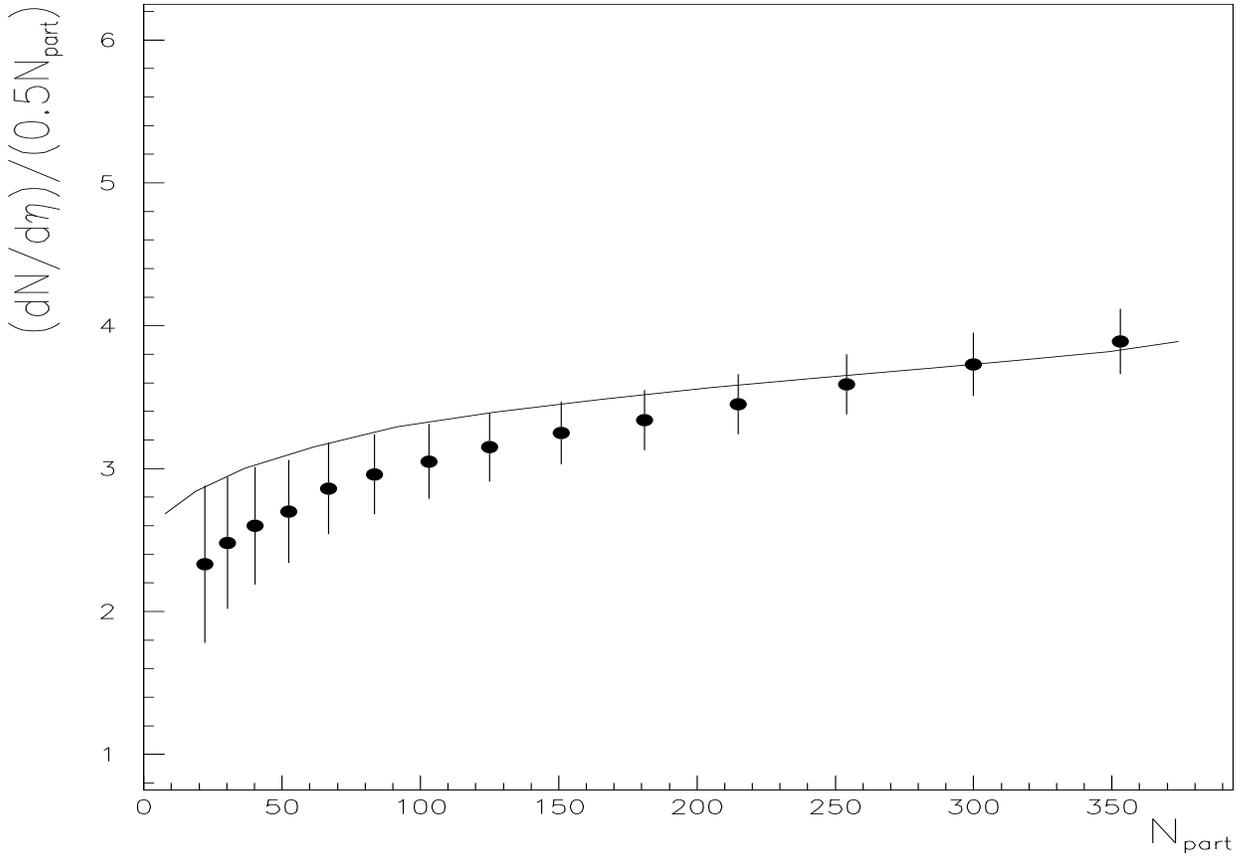}
\vskip -1.5cm
\caption{Multiplicity of charged particles per participant pairs, 
versus $N_{part}$, computed from (\ref{13e}) are compared to 
experimental data from
PHENIX {\protect\cite{15r}}.}
\end{figure}

\newpage

\begin{figure}
\centering\leavevmode
\vskip 4.5cm
\epsfxsize=6in\epsfysize=6in\epsffile{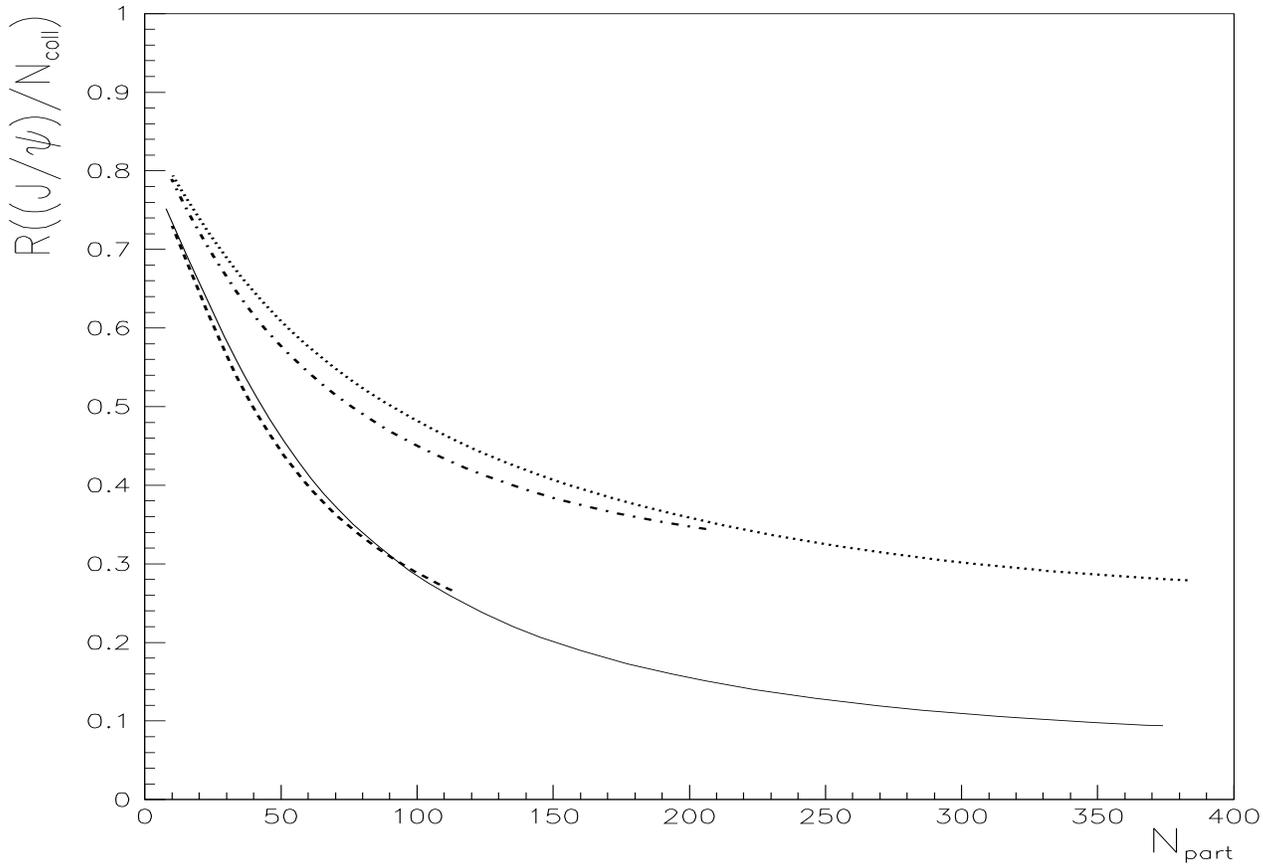}
\vskip -1.5cm
\caption{$R_{AB}^{J/\psi}(b)$ for $AuAu$ collisions at $\sqrt{s}=200$ GeV
(full curve), $CuCu$ collisions at $\sqrt{s}=200$ GeV (dashed curve),
$PbPb$ at $p_{lab}=158$ GeV/c (dotted curve)
and $InIn$ at $p_{lab}=158$ GeV/c (dashed-dotted curve).
All the results have been obtained with $\sigma_{co} = 0.65$~mb and
$\sigma_{abs} = 4.5$~mb. The normalization, the same for all four 
curves, is arbitrary. It corresponds to taking $dN_{pp}^{J/\psi}/dy = 
1$ in eq. (\ref{1e}).}
\end{figure}

\newpage

\begin{figure}
\centering\leavevmode
\vskip 4.5cm
\epsfxsize=6in\epsfysize=6in\epsffile{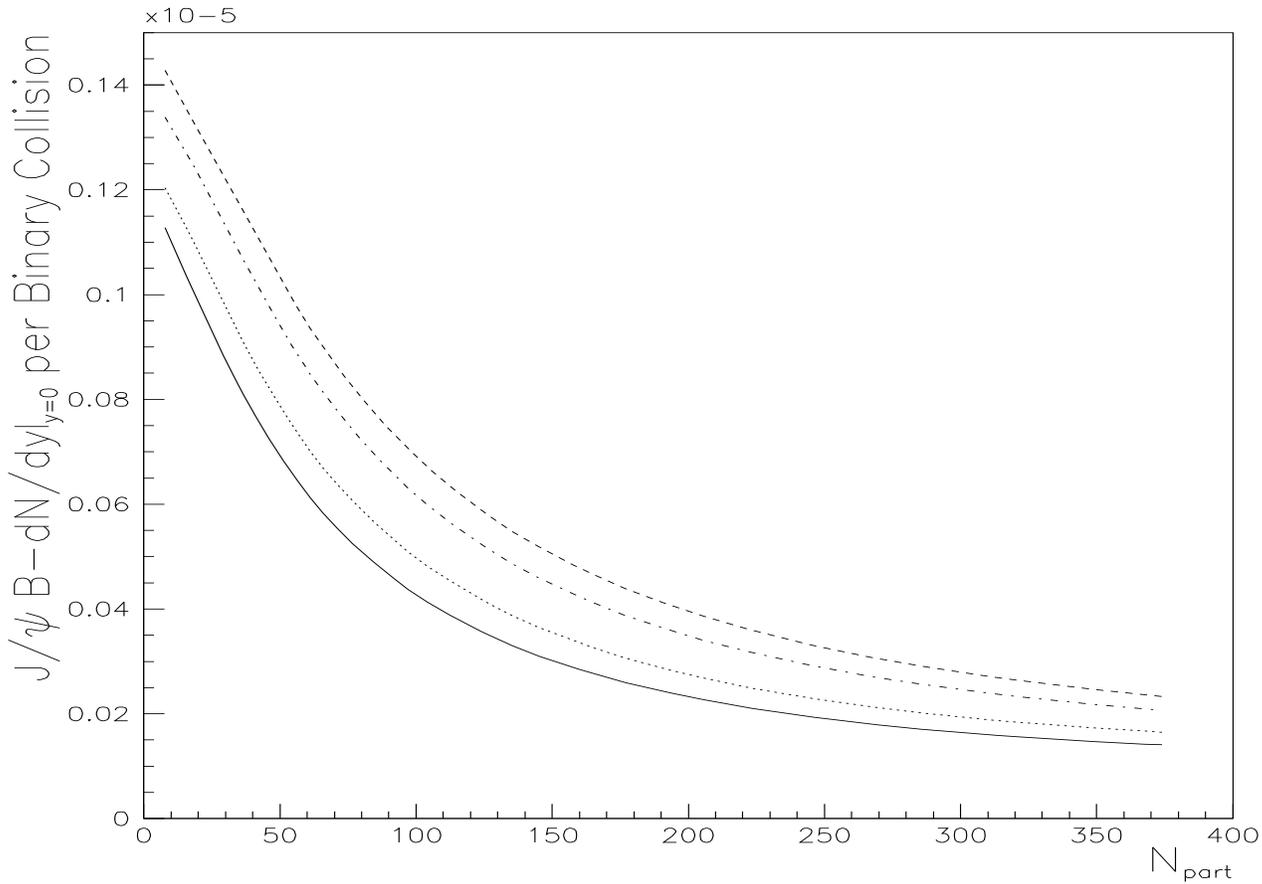}
\vskip -1.5cm
\caption{$R_{AB}^{J/\psi}(b)$ for $AuAu$ collisions at $\sqrt{s}=200$
GeV multiplied by the dilepton branching ratio, normalized to the 
measured value in $pp$ collisions {\protect\cite{16r}}. From up to 
down:
result with $\sigma_{co} = 0.65$~mb and
$\sigma_{abs} = 0$~mb (dashed curve), result with $\sigma_{co} = 0.65$~mb and
$\sigma_{abs} = 1$~mb (dotted-dashed curve), result with $\sigma_{co} 
= 0.65$~mb and
$\sigma_{abs} =3$~mb (dotted curve) and result with $\sigma_{co} = 0.65$~mb and
$\sigma_{abs} =4.5$~mb (full curve). }
\end{figure}

\end{document}